# Observation of Intrinsic and LED Light-Enhanced Memristor Performance in In-Plane Ferroelectric NbOI$_2$

*Zheng Hao,[1,2] Gaolei Zhao,[1] Haoran Li,[3] Jichang Zhang,[1] Jiabin Liu,[1] Fanyi Kong,[1] Konstantin Kozadaev,[4] Yongjiang Li,[3] Xue Han,[1,2] Hong Li,[1] Huolin Huang,[1] Changsen Sun,[1] Alexei Tolstik,[5] Andrey Novitsky,[4] Lujun Pan,[6] and Dawei Li[1,2,]\**

[1] School of Optoelectronic Engineering and Instrumentation Science, Dalian University of Technology, Dalian, Liaoning 116024, China

[2] Dalian University of Technology and Belarusian State University Joint Institute, Dalian University of Technology, Dalian, Liaoning 116024, China

[3] School of Biomedical Engineering, Faculty of Medicine, Dalian University of Technology, Dalian, 116024, China

[4] Physical Optics and Applied Informatics Department, Belarusian State University, Minsk, 220030, Belarus

[5] Laser Physics and Spectroscopy Department, Belarusian State University, Minsk, 220030, Belarus

[6] School of Physics, Dalian University of Technology, Dalian, Liaoning 116024, China

\* Address correspondence to: dwli@dlut.edu.cn

**ABSTRACT**: Two-dimensional (2D) layered ferroelectrics, as an emerging area of research, have attracted extensive attention, while memristors based on new 2D ferroelectric materials have yet to be fully explored, thereby limiting their applications in modern nanoelectronics. In this work, we report the observation of intrinsic memristive behavior in a newly discovered 2D in-plane ferroelectric material, NbOI$_2$, and the giant enhancement of the memristive performance using LED visible light. The results show that NbOI$_2$ exhibits intrinsically strong memristive response with a current on/off ratio of up to $10^4$ and stable switching cycles (≥20), which is independent of back-gate voltage. Under LED visible light illumination, the current on/off ratio in NbOI$_2$ is over one order of magnitude higher than that without light, meanwhile, the coercive field is significantly reduced to less than ~1.22 kV/cm, much lower than other 2D ferroelectric material-based memristors. Interestingly, both the intrinsic and the light-enhanced resistive switching phenomena only occur along the in-plane *b*-axis direction, indicating that the memristive behavior in NbOI$_2$ is driven by electric field-induced and optical field-enhanced ferroelectric polarization switching mechanism, as evidenced by a combined orientation-dependent electrical/optoelectrical measurement and sweep cycle-induced structural evolution analysis. Our study not only provides a materials strategy based on new 2D ferroelectrics for designing memristor applications, but also offers a simple optical method to enhance its performance, paving the path for its implementation in novel nanoelectronics and optoelectronics.

**INTRODUCTION**

The rapid advancement of Internet of Things and neuromorphic computing technologies has increasingly exposed the limitations of conventional von Neumann architecture, thereby driving significant research interest in developing novel non-volatile electronic devices.[1-2] Memristors, a kind of resistive device with an inherent memory effect,[3-4] have emerged as promising candidates for realizing in-memory computing architectures, due to their structural simplicity, low power consumption, and CMOS compatibility.[5-6] However, memristors based on traditional bulk materials (e.g., HfO$_2$,[7] SiO$_2$,[8] BiFeO$_3$,[9] TaO$_x$,[10] etc) face critical challenges, including stochastic switching mechanisms, insufficient cycling stability, and device size limitation,[11-12] thus hindering their applications in high-performance neuromorphic hardware systems. In recent years, two-dimensional (2D) van der walls (vdW) materials have attracted much attention due to their atomically thin layer thickness, excellent electrical, optical and mechanical properties.[13-14] In addition, significant progress has been made in memristors based on 2D materials,[15-16] including *h*-BN,[17-18] graphene,[19-20] 2D semiconductors,[21] and transition metal chalcogenides.[22-23] However, many new 2D materials and functional devices need to be further explored.

Recently, 2D vdW ferroelectrics, as an emerging area of research, have attracted extensive attention as they can retain robust ferroelectricity down to the monolayer limit,[24-25] and show great potential applications in nanoelectronics and optoelectronics.[26] Typical 2D ferroelectric materials include $CuInP_2S_6$ with out-of-plane polarization,[27-28] α-$In_2Se_3$ with intercorrelated out-of-plane and in-plane polarization,[29-30] and $NbOX_2$ (X = Cl, I) with in-plane polarization.[31-32] According to the previous research,[33] 2D ferroelectrics are also promising as active materials for high-performance memristive devices. For example, Li et al. successfully demonstrated a memristor based on 2D ferroelectric $CuInP_2S_6$ with out-of-plane polarization, which shows good current on/off ratio (~6×$10^3$).[34] More recently, Jeon et al. fabricated a planar memristor using 2D ferroelectric α-$In_2Se_3$ with intercorrelated out-of-plane and in-plane polarization.[35] Although a much larger current on/off ratio can be achieved in this device, the complicated Au/α-$In_2Se_3$/$SiO_2$/Au junction structures are required. As far as we know, research on the intrinsic memristive properties of 2D ferroelectric materials with in-plane polarization and how to enhance these properties have not been fully explored to date.[36]

In this work, we report the observation of stable and strong memristive properties in in-plane vdW ferroelectric $NbOI_2$ and the giant enhancement of memristive performance in $NbOI_2$ by an external optical field. We have fabricated two-terminal planer devices and three-terminal field-effect transistor (FET) devices using single-crystal multilayer $NbOI_2$ as channel material and $SiO_2$ as back gate. Experimental results show that 2D $NbOI_2$ devices exhibit intrinsically strong memristive effect with a current on/off ratio of up to $10^4$ and stable switching cycles (≥20), which is independent of back-gate voltage. More importantly, under illumination by using red light-emission diode (LED) as a light source, the devices show enhanced memristive performance, where the current on/off ratio is over one order of magnitude higher than that without LED light, meanwhile the coercive voltage (or coercive field) is significantly reduced to less than 0.8 V (or ~1.22 kV/cm). These memristive behaviors only occur when the current signals are detected along the ferroelectric polarization (or *b*-axis) direction, indicating that the memristive behavior in $NbOI_2$ is driven by electric field-induced and optical field-enhanced ferroelectric switching effect, as evidenced by a combination of direction-dependent electrical/optoelectrical measurement and sweep cycle-induced structure evolution analysis. Our study not only points to a materials strategy based on 2D in-plane vdW ferroelectrics for designing memristive device applications, but also offers a simple optical method to significantly enhance its performance.

**RESULTS AND DISCUSSION**

$NbOI_2$ is a layered material with monoclinic structure within space group C2 (Figure 1a).[37] In detail, the framework of $NbOI_2$ is composed of a 2D network of octahedra, which is interconnected via I-I edges along the *c*-axis, while oxygen atoms bridge adjacent units along the *b*-axis (Figure 1b and 1c). It is noted that Nb atoms are displaced off-center within the octahedra along the *b*-axis direction, leading to a Peierls distortion. This structural distortion establishes that *b*-axis crystal orientation corresponds to polar direction, which is responsible for the intrinsic ferroelectric polarization in $NbOI_2$. Due to the weak interlayer vdW interaction, $NbOI_2$ thin layers can be easily achieved by mechanical exfoliation from the bulk single crystals. Figure 1d shows the optical image of a mechanically exfoliated $NbOI_2$ flake transferred onto the $SiO_2$/Si substrate. The thickness of this $NbOI_2$ sample is determined by atomic force microscopy (AFM) measurement, which is approximately 100 nm (Supporting Information).

To identify the in-plane *c*-axis and *b*-axis in $NbOI_2$, we performed a polarized optical imaging measurement (Figure 1e).[38] When the sample in Figure 1d is placed in a parallel polarized optical path, we find that the brightness becomes brighter as the sample is rotated by $\theta = 90°$ (Figure 1f and 1g) and then becomes weaker by a further rotation (Supporting Information). Figure 1h shows the polar plot of parallel polarized reflected light intensity as a function of sample rotation angle, which exhibits a distinct two-lobed shape. It has been shown that the maximum light absorption coefficient could be achieved when the visible light polarization is parallel to the *c*-axis of NbOI$_2$.[39] Therefore, the maximum and minimum value directions in Figure 1h correspond to the *b*-axis and *c*-axis, respectively.

Figure 1i shows the Raman spectrum of the NbOI$_2$ sample in Figure 1d, from which five distinct vibrational peaks can be clearly observed. The four characteristic peaks located at ~104 cm$^{-1}$, ~207 cm$^{-1}$, ~271 cm$^{-1}$, ~608 cm$^{-1}$ are originated from $A_g$-like mode, while a relatively weak peak at ~510 cm$^{-1}$ is originated from $B_g$-like symmetry mode, consistent with the previously reported results.[39] To further confirm the in-plane crystal orientation, we performed angle-resolved Raman measurement in parallel polarized configuration (Figure 1j). From the angular-dependent Raman peak intensity analysis (Figure 1k and Supporting Information), we observe that all the peaks for $A_g$-like mode exhibit a two-lobed shape, while the peak for $B_g$-like mode displays a four-lobed shape. Based on the above analysis, the in-plane (*b*-axis and *c*-axis) crystal orientation of NbOI$_2$ is determined,[39] in agreement with that by parallel polarized optical imaging method (Figure 1h). In addition, these characterizations demonstrate that NbOI$_2$ shows strong in-plane optical and structural anisotropy, which is also expected to exhibit unique electrical and optoelectrical behaviors.

To probe the electrical properties of NbOI$_2$, we fabricate three-terminal FET devices using multilayer NbOI$_2$ with controlled crystal orientation as channel material and SiO$_2$ as back gate (see Method). We first investigate the electrical transport of NbOI$_2$ along the *b*-axis (or ferroelectric polarization) direction (Figure 2a). Figure 2b shows the optical image of a multilayer NbOI$_2$ device (D1), with its *b*-axis perpendicular to two Au electrodes, as confirmed by parallel polarized light intensity analysis (inset). Figure 2c shows the source-drain current-voltage ($I_d$-$V_d$) characteristic of the device at a maximum sweep voltage ($V_{d,max}$) of ±15 V, which exhibits a hysteresis phenomenon. The arrows in Figure 2c point to the $V_d$ sweep direction. It can be seen that, as the $V_d$ sweeps from 0 V to +15 V (stage 1), the device is switched from the high-resistance state (HRS) to the low-resistance state (LRS), and retains in the LRS when the $V_d$ is decreased back to 0 V (stage 2). Similarly, as the $V_d$ sweeps from 0 V to -15 V (stage 3), the device is switched from HRS to LRS, and keeps in the LRS until the $V_d$ is decreased back to 0 V (stage 4). It is suggested that NbOI$_2$ along *b*-axis behaves as a typical memristor. The similar resistive switching behavior has been demonstrated in more NbOI$_2$ devices (Supporting Information).

Next, we compare the $I_d$–$V_d$ curves of the NbOI$_2$ device taken at different $V_{d,max}$ (Figure 2d). No significant memristive behavior is observed when $|V_{d,max}| < 5$ V. When $|V_{d,max}| \geq 5$ V, the memristive behavior is detected, and progressively intensified with increasing $V_{d,max}$. We then calculate the current on/off ratio ($I_{on}/I_{off}$ or $R_{LRS}/R_{HRS}$) of the device for both positive and negative bias directions at each $V_{d,max}$ (Figure 2e), where a clear positive correlation between $I_{on}/I_{off}$ and $V_{d,max}$ is observed. The maximum current on/off ratio can exceed $1\times10^3$, comparable to or even higher than other 2D material-based memristors.[27, 36, 40-42] We also investigated the switching endurance characteristic of the NbOI$_2$ device. Figure 2f shows the $I_d$–$V_d$ curves of the NbOI$_2$ device (D1) during 20 sweep cycles, which exhibits a good reproducibility of the current switching behavior. Figure 2g shows the statistical distribution of resistance in the HRS and LRS extracted from Figure 2f, demonstrating the stable resistance level with increasing the sweep cycles. The average resistance value in the HRS and LRS is $3.5\times10^{11}$ Ω and $1\times10^8$ Ω, respectively, corresponding to an average $R_{LRS}/R_{HRS}$ ratio of $3.5\times10^3$. Such stable switching behavior with high $R_{LRS}/R_{HRS}$ ratio (~$0.9\times10^4$) is observed in other NbOI$_2$ devices (D2, Supporting Information). Furthermore, the effect of gate voltage ($V_g$) on the memristive behavior of NbOI$_2$ is studied, as shown in Figure S4g. It is found that the electrical transport properties of NbOI$_2$ remain unchanged by varying $V_g$ from -20 V to +20 V, indicating of gate-independent memristive behavior characteristics.

It is noted that, under ambient condition without external stimuli, NbOI$_2$ devices exhibit memristive effect primarily at relatively high sweep voltage ($|V_{d,max}| > 5$ V), and a good memristive performance is usually obtained at $|V_{d,max}| \geq 10$ V. To improve its memristive performance, we implement LED visible light (~610 nm) illumination in $NbOI_2$ device during the electrical measurement (Figure 3a). Figure 3b presents the optical image of a multilayer $NbOI_2$ device (D3) with its *b*-axis perpendicular to two Au electrodes. When we measure the $I_d$–$V_d$ curve of D3 in dark condition (Figure 3c), the device exhibits negligible (obvious) memristive behavior at $|V_{d,max}| < 6$ V ($\geq 10$ V). Under 610 nm LED light exposure condition (Figure 3d), the device demonstrates pronounced memristive characteristics with two distinct improvements: (1) The resistance state switching occurs at significantly reduced operational voltage ($\leq 0.8$ V), which is much lower than that in dark condition. (2) The current on/off ratio exhibits substantial enhancement compared with that measured in dark condition. For instance, at $V_{d,max} = \pm 10$ V, the current on/off ratio under LED light illumination is over one order of the magnitude higher than that in the dark. Figure 3e shows the $I_d$–$V_d$ curves of D3 taken immediately after the LED light turned off, where a progressive attenuation in the current response is observed, indicative of a persistent polarization relaxation process in the $NbOI_2$ device. After four sweep cycles, the memristive effect completely disappears at $|V_{d,max}| < 6$ V, while significant memristive behavior retains at $|V_{d,max}| \geq 10$ V, similar to that taken in initial dark condition.

To understand the above observations, we comparatively investigate the electrical transport characteristics of $NbOI_2$ along different crystal orientations (Figure 4). Figure 4a and 4b show the schematic and optical image of a two-terminal multilayer $NbOI_2$ device (D4), with its in-plane *c*-axis perpendicular to two Au electrodes. As shown in Figure 4c (black curve), D4 exhibits distinct electrical behavior compared to the device along the *b*-axis (Figure 2), where no memristive behavior is observed. Under LED light exposure condition (red curves), D4 demonstrates the increased source-drain current signal by about 10-fold, but fails to induce any resistive switching behavior. To investigate the electrical transport of $NbOI_2$ along the out-of-plane (*a*-axis) direction, a vertical device architecture is constructed by using multilayer graphene as both top and bottom electrodes and $NbOI_2$ as channel material (Figure 4d). Figure 4e shows the optical image of a typical graphene/NbOI$_2$/graphene vertical device (D5), where the orange and red dashed lines indicate graphene and NbOI$_2$, respectively. Similar to that along the *c*-axis, no memristive behavior is observed in vertical NbOI$_2$ device over the entire $V_d$ scanning range, whether in the dark or light illumination condition (Figure 4f and Supporting Information).

Based on the above analyses, we conclude that both the pristine (Figure 2) and the light-enhanced (Figure 3) resistive switching behavior in NbOI$_2$ is highly dependent on the crystal orientations, which is only realized along the *b*-axis direction, completely different from the previously reported experimental result.[43] It is indicated that, in our case, the ferroelectric polarization switching plays a critical role in the observed memristive behavior, namely, NbOI$_2$ can work as a ferroelectric memristor. Therefore, comparison of electrical transport in NbOI$_2$ without and with light illumination (Figure 3) demonstrates that external optical field may enhance the ferroelectric-driven memristor performance and significantly reduce the coercive voltage ($V_C$). Under 610 nm light exposure, $V_c$ is measured to be ~0.8 V, corresponding to a coercive field of ~1.22 kV/cm, which is much lower than that of some other 2D ferroelectric materials.[35-36, 44]

To understand the physical mechanism of memristive behavior observed in NbOI$_2$, we further analyze the current switching characteristic of NbOI$_2$ along the *b*-axis (Figure 4g), which can be explained by the asymmetric modulation of the Schottky barrier via the ferroelectric polarization charges (Figure 4h).[45] The yellow arrows in Figure 4h indicate the direction of ferroelectric polarization ($\vec{P}$) in NbOI$_2$ device. As $|V_d| < V_C$, the ferroelectric dipoles of NbOI$_2$ are randomly orientated, leading to no net polarization charges at either source or drain end. Under this condition, symmetric Schottky barriers are formed at two ends (State 1), with the barrier height determined by the intrinsic band edge energies and Fermi levels of NbOI$_2$ and Au. As a positive $V_d$ ($|V_d| > V_C$) is applied (in sweep 1 and sweep 2 of Figure 4g), $\vec{P}$ points right direction, leading to the accumulation of positive (negative) charges at the source (drain) end. Under this condition, the Schottky barrier decreases at the source end (State 2), and the current signal significantly increases, resulting in the LRS. Similarly, as a negative $V_d$ ($|V_d| > V_c$) is applied (in sweep 3 and sweep 4 of Figure 4g), positive (negative) polarization charges are generated at the drain (source) end. Under this condition, the height barrier at the drain end decreases (State 3), and the current signal also significantly increases, forming the LRS. Hence, the bias voltage-controlled ferroelectric polarization switching mechanism governs the different resistance states in the NbOI$_2$ memristor devices.

The previous studies have shown that the resistive switching process in memristors is usually accompanied by structural change,[46-47] especially after multiple switching cycles. Next, we investigated the ferroelectric switching induced structural evolution in NbOI$_2$, which may deepen our understanding of its working principle. Here, we fabricated a two-terminal NbOI$_2$ device along the *b*-axis (D6) (Figure 5a), with the bottom half of the channel connected with two Au electrodes and the top half unconnected. The electrical transport measurement ($V_{d,max} = \pm 20$ V) confirms distinct memristive behavior in D6 (Figure 5b). *In-situ* optical imaging was collected after different switching cycles, as shown in Figure 5c-5e. It is clearly observed that several darkened line patterns emerge along the current path direction (*b*-axis) at the bottom half of the channel, and progressively expand with prolonged testing duration. As expected, this phenomenon does not occur at the top half of the channel, indicative of no structural change. To investigate the nature of these line patterns, we performed AFM imaging analysis of D6 after 80 switching cycles (Figures 5f). High-magnified AFM imaging (Figure 5g and 5h) shows that the line patterns exhibit slight protrusion or depression in topography, revealing the sweep cycle-induced localized morphological change characteristics in NbOI$_2$ during the electrical transport measurement.

To confirm the structural evolution in NbOI$_2$, we also performed polarized optical imaging study. Figure 5i compares the polar plots of parallel polarized light intensity as a function of sample rotation angle for three local regions in Figure 5e [pristine region (P1), line patterned regions (P2 and P3)]. The pristine region (P1) maintains the characteristic two-lobed polarization pattern, demonstrating the preserved in-plane structural anisotropy. In contrast, both P2 and P3 regions display significantly reduced in-plane optical anisotropy, indicating of structural modification in the line patterns. To further confirm the structural evolution, we have conducted Raman spectroscopy analysis (Figure 5j-5l). Figure 5j compares the Raman spectra of three representative regions, P1, P2, and P3. Both P2 and P3 regions exhibit weaker Raman signal than P1 region, which is more clearly distinguished by Raman mapping characterization (Figure 5k). Notably, the characteristic Raman peaks of $A_g$-like mode (at ~272 cm$^{-1}$ and ~609 cm$^{-1}$) for modified regions show a distinct blue shift (Figure 5j and 5l), further corroborating the local structural change in NbOI$_2$. Hence, the correlation between structural evolution and switching cycle yields strong support to the idea that the observed memristive behavior in NbOI$_2$ originates from electric field-induced polar structural change in specific channel regions.

## CONCLUSIONS

In summary, we have systemically investigated the electrical and optoelectrical properties in 2D layered ferroelectric NbOI$_2$, which are highly dependent on the crystal orientations. We found that NbOI$_2$ exhibits intrinsically strong memristive behavior along the in-plane *b*-axis (ferroelectric polarization) direction, with a current on/off ratio of up to 10$^4$ and stable switching cycles of over 20. However, no memristive effect is detected along other crystal directions, including in-plane *c*-axis and out-of-plane *a*-axis. By using the visible LED as a light source, the current on/off ratio in NbOI$_2$ memristor can be enhanced by over one order of magnitude, and the coercive field is dramatically reduced to less than ~1.22 kV/cm, lower than most of the 2D ferroelectric-based memristors. In addition, a ferroelectric polarization switching model is proposed, which can well explain the working principle of the memristive behavior in NbOI$_2$. Overall, our study demonstrates that NbOI$_2$ as a newly discovered 2D vdW ferroelectric material shows great promise for novel low-power, non-volatile device applications.

## METHODS

**Sample preparation**. The NbOI$_2$ flakes are mechanically exfoliated from their bulk single crystal (Nanjing MKNANO Tech., China) and then transferred onto the SiO$_2$/Si substrates, with their layer thicknesses determined by AFM measurement. The in-plane crystalline orientations of NbOI$_2$ flakes are identified by combining parallel polarized optical imaging and parallel polarized Raman spectroscopy measurements.[38]

**Device fabrication.** The NbOI$_2$ devices are fabricated by the standard photolithography process. In brief, Au/Ti (15 nm/5 nm) electrodes are prepatterned on the SiO$_2$/Si substrate, and then NbOI$_2$ flakes on a gel film are transferred to the top of the prepatterned Au/Ti electrodes using a dry transfer technique.[48-49] In addition, we have fabricated the graphene/NbOI$_2$/graphene vertical heterostructure devices by dry transfer technique, through which out-of-plane electrical measurement can be performed.

**Characterizations**. The morphology and thickness measurements are carried out in an AFM system (MFP-3D Origin, Oxford, UK). The optical characterizations are conducted using an optical microscopy (LV150N, Nikon, Japan) with a tungsten halogen lamp as the light source and in a micro-Raman system (Renishaw inVia, UK). Raman signals are collected by focusing a 532 nm laser onto the sample surface through a 50× objective with an exposure time of 10 s. The electrical measurements are performed using a semiconductor parameter analyzer (B1500A, Keysight, USA) in the dark or under a 610 nm LED light (Oeabt LEDOTB-100) illumination.

## ACKNOWLEDGEMENTS

This work was supported by the National Natural Science Foundation of China (Grant No. 12274051), the Natural Science Foundation of Liaoning Province (Grant No. 2024-MSBA-06), the Liaoning Province Xingliao Talents Plan Project (Grant No. XLYC2403069), the Chunhui Project Foundation of the Education Department of China (Grant No. HZKY20220423), the Fundamental Research Funds for the Central Universities (Grant Nos. DUT24RC(3)060, DUT22ZK109), and the National Key Research and Development Program of China (Grant No. 2024YFE0213500).

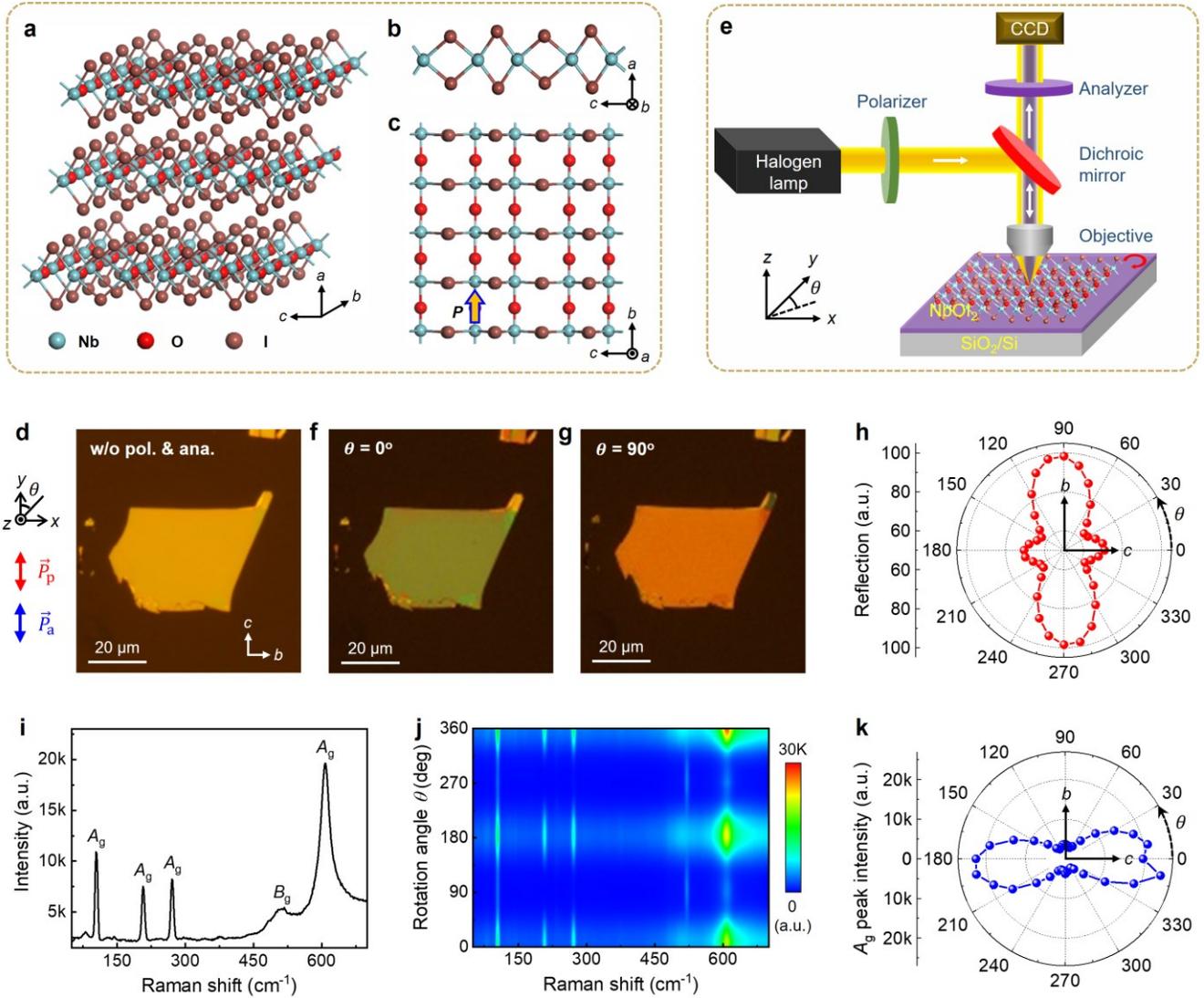

**Figure 1.** Determination of polar axis in layered ferroelectric NbOI$_2$. (a-c) Schematics of crystal structure of 2D NbOI$_2$: (a) 3D view, (b) side view, (c) top view. The in-plane *b*-axis orientation corresponds to the direction with polarity ($\vec{P}$). (d) Optical images of an exfoliated thick-layer NbOI$_2$ flake taken without polarizer and analyzer. (e) Schematic of polarized optical imaging setup. (f and g) Parallel polarized optical images of the same sample in (d) taken at different rotation angles $\theta$. (h) Polar plot of parallel polarized light intensity for NbOI$_2$ in (d) as a function of sample rotation angle. (i) Raman spectrum for the same sample in (e). (j) Angular resolved Raman spectra for NbOI$_2$ in (d) in parallel polarized configuration. (k) Polar plot of parallel polarized Raman peak intensity at $A_g$-band as a function of sample rotation angle.

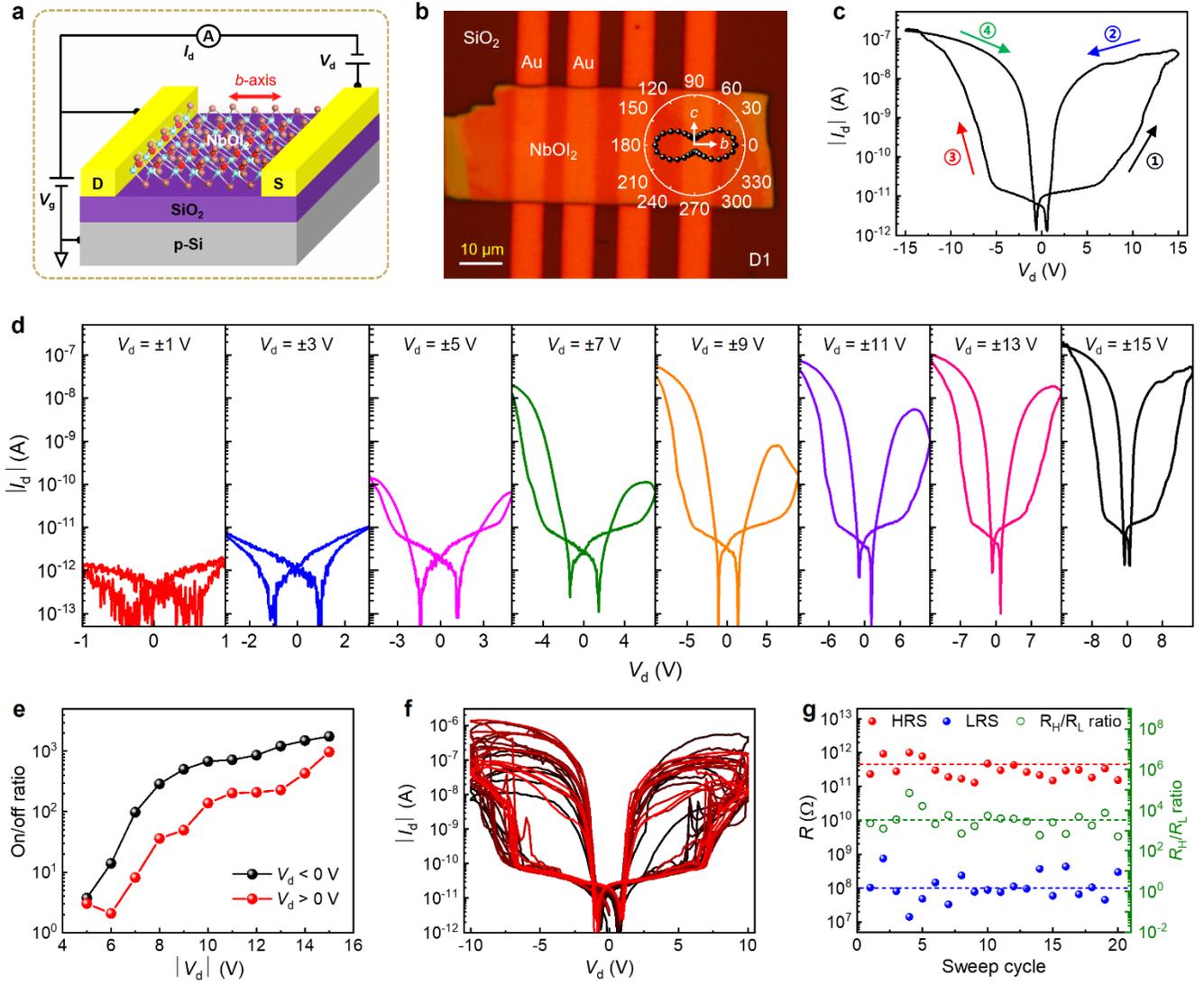

**Figure 2.** Memristive effect in layered ferroelectric NbOI$_2$. (a) Schematic of NbOI$_2$ FET device using SiO$_2$ as the bottom gate, where the polarization direction (*b*-axis) of NbOI$_2$ is perpendicular to two Au electrodes. (b) Optical image of a multilayer NbOI$_2$ device. Inset: Polar plot of parallel polarized light intensity as a function of sample rotation angle. (c) Output ($I_d$−$V_d$) characteristic ($V_{d,max}$ = ±15 V) of the device in (b) represented in logarithmic coordinate system. (d) $I_d$−$V_d$ curves of the NbOI$_2$ device taken at different $V_{d,max}$. (d) Relationship between current $I_{on}/I_{off}$ ratio and $V_d$. (f) $I_d$−$V_d$ curves of the NbOI$_2$ device during 20 sweep cycles. (g) Statistical distribution of (left) resistances in the HRS ($R_H$, red dots) and LRS ($R_L$, blue dots) and (right) $R_H/R_L$ ratio (green dots) during 20 cycle process in (f). Dashed lines represent the averaged $R_H$, $R_L$, and $R_H/R_L$ ratio during 20 cycle process.

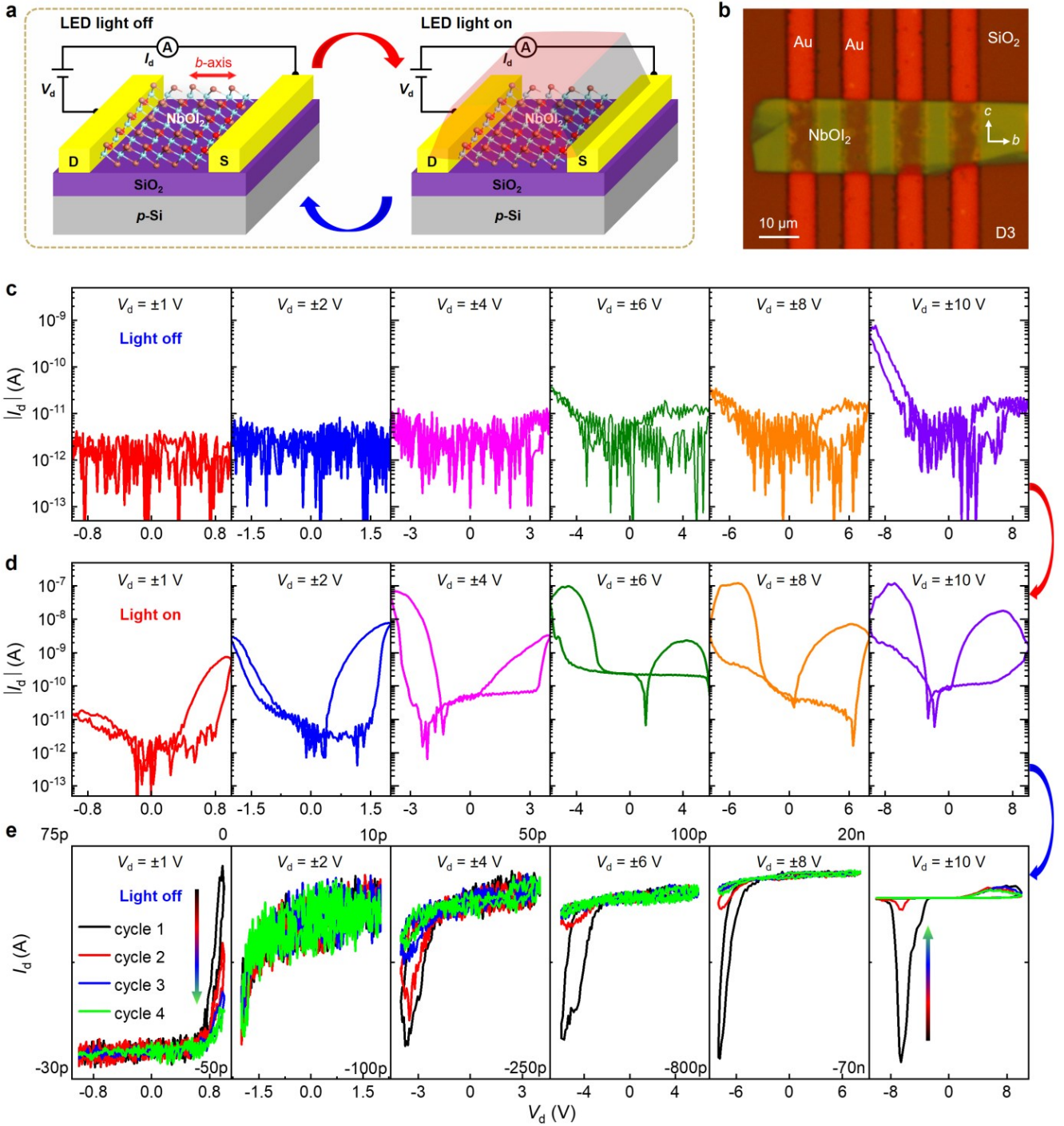

**Figure 3.** LED visible light-enhanced memristive performance in NbOI$_2$. (a) Schematic of NbOI$_2$ device without (left) and with (right) light illumination. (b) Optical image of a NbOI$_2$ device, with the *b*-axis of NbOI$_2$ perpendicular to two Au electrodes. (c and d) $I_d$−$V_d$ curves of the device in (b) taken in the dark (c) and under 610-nm LED light exposure (d) at different $V_{d,max}$. (e) $I_d$−$V_d$ curves of the device in (b) taken when the LED light turned off again, under continuous voltage scanning.

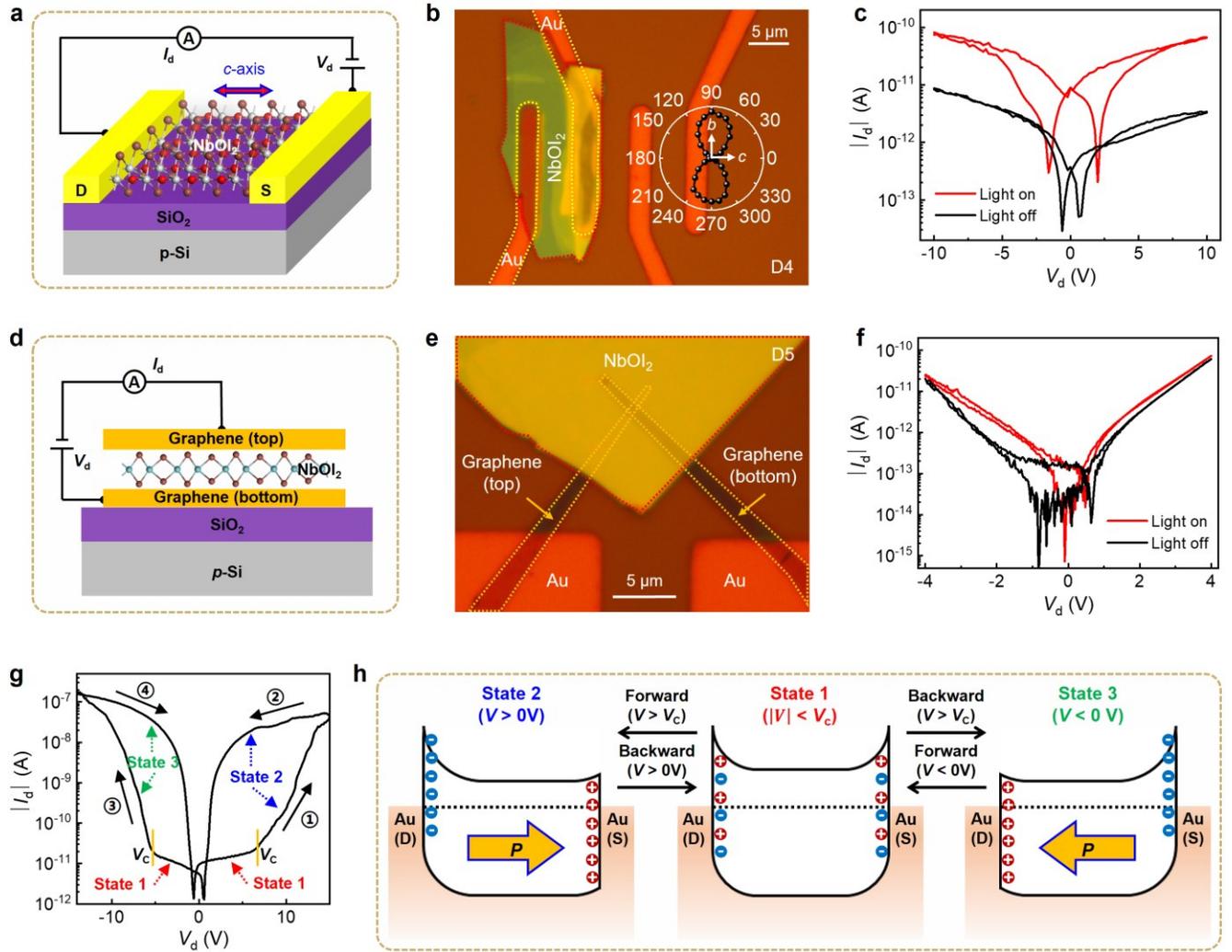

**Figure 4.** Evidence for ferroelectric polarization-driven memristive effect in NbOI$_2$. (a) Schematic of NbOI$_2$ device, where the *c*-axis of NbOI$_2$ is perpendicular to two Au electrodes. (b) Optical image of a multilayer NbOI$_2$ device, with its *c*-axis parallel to channel direction, as evidenced by parallel polarized light intensity analysis (inset). (c) $I_d$–$V_d$ characteristic curves ($V_{d,max}$ = 10 V) of the device in (b) taken in the dark and under 610-nm LED light exposure. (d) Schematic of NbOI$_2$ vertical device using graphene as the top and bottom electrodes. (e) Optical image of a graphene/NbOI$_2$/graphene vertical device. The red and orange dotted lines mark the boundaries of NbOI$_2$ and graphene, respectively. (f) $I_d$–$V_d$ characteristic curves ($V_{d,max}$ = 4 V) of the device in (e) taken in the dark and under 610-nm LED light exposure. (g) Typical $I_d$–$V_d$ curve of the NbOI$_2$ device along the polar direction, indicating (h) ferroelectric polarization switching-driven memristive behavior.

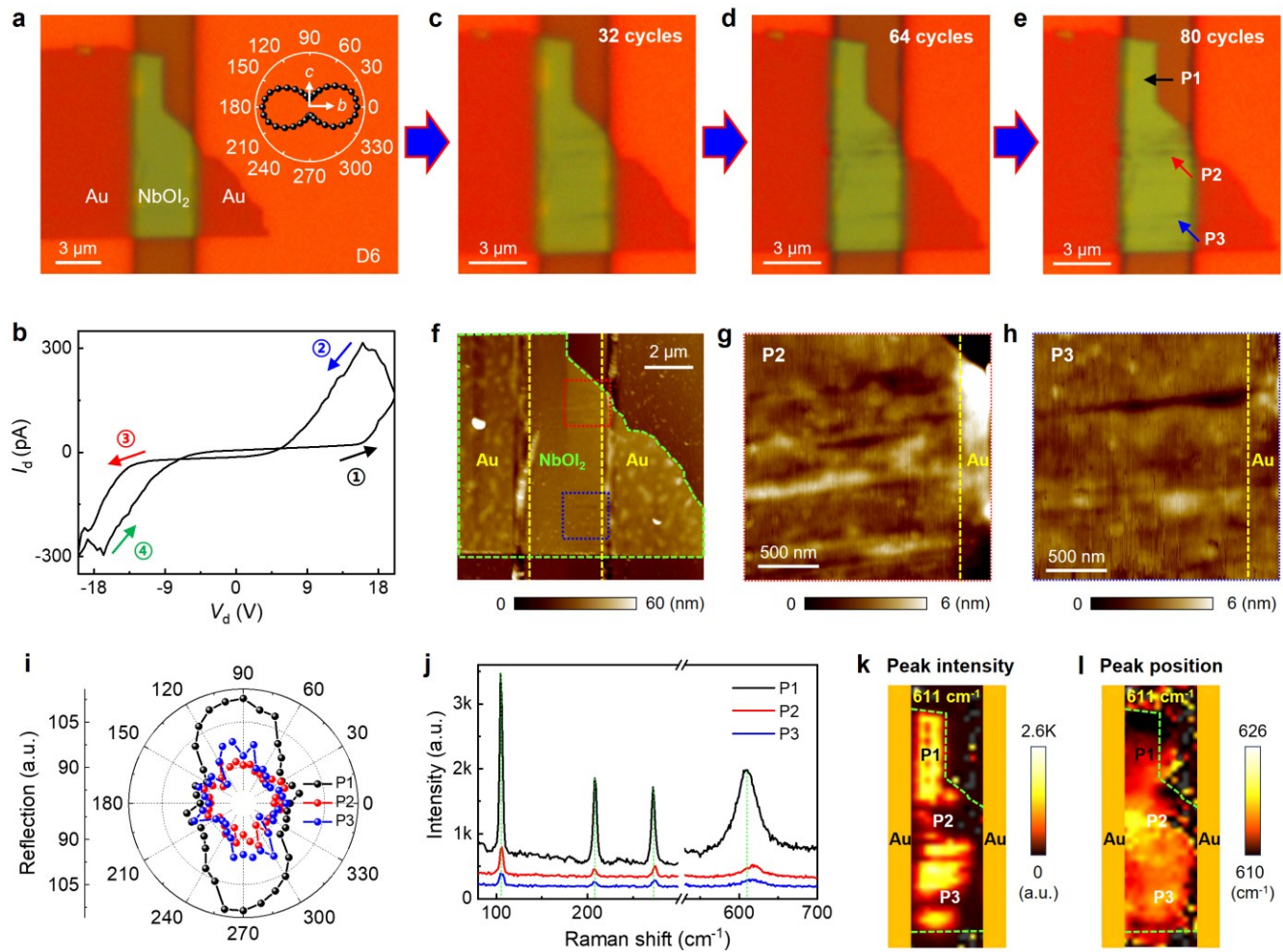

**Figure 5.** Observation of ferroelectric polarization switching-induced structural change in $NbOI_2$. (a) Optical image of a multilayer $NbOI_2$ device, with its *b*-axis parallel to channel direction, as evidenced by parallel polarized light intensity analysis (inset). (b) Typical $I_d$–$V_d$ curve for the device in (a). (c-e) Optical images of the $NbOI_2$ device in (a) after different switching cycle processes: (c) 32 cycles, (d) 64 cycles, and (e) 80 cycles. (f) AFM topography image of the device in (a) after 80 switching cycles. (g and h) High-magnified AFM images in (f) at red boxed area (g) and blue boxed aera (h). The green and yellow dashed lines in (f-h) mark the edges of $NbOI_2$ and Au electrode, respectively. (i) Polar plots of parallel polarized light intensity for three local regions (P1, P2, and P3) in (e) as a function of sample rotation angle. (j) Raman spectra for three local regions (P1, P2, and P3) in (e). (k, l) Raman mapping of (k) 611 cm$^{-1}$ peak intensity and (l) 611 cm$^{-1}$ peak position taken from $NbOI_2$ device in (e).